\documentclass[preprint,showpacs,amsmath,amssymb,prb,floatfix]{revtex4}

\usepackage{graphicx}
\usepackage{dcolumn}
\usepackage{bm}
\input epsf.sty

\begin{document}

\title{Ising metamagnets in thin film geometry: equilibrium properties}

\author{Yen-Liang Chou and Michel Pleimling}
\affiliation{
Department of Physics,
Virginia Tech,
Blacksburg, VA 24061-0435, USA}

\date{\today}

\begin{abstract}
Artificial antiferromagnets and synthetic metamagnets have attracted much attention recently due to their
potential for many different applications. Under some simplifying assumptions these systems can be modeled
by thin Ising metamagnetic films. In this paper we study, using both the Wang/Landau scheme and importance
sampling Monte Carlo simulations, the equilibrium properties of these films. On the one hand we discuss the microcanonical
density of states and its prominent features. On the other we analyze canonically various global and layer
quantities. We obtain the phase diagram of thin Ising metamagnets as a function of temperature and external
magnetic field. Whereas the phase diagram of the bulk system only exhibits one phase transition between the
antiferromagnetic and paramagnetic phases, the phase diagram of thin Ising metamagnets includes an additional
intermediate phase where one of the surface layers has aligned itself with the direction of the applied
magnetic field. This additional phase transition is discontinuous and ends in a critical end point. Consequently,
it is possible to gradually go from the antiferromagnetic phase to the intermediate phase without passing through 
a phase transition.
\end{abstract}

\pacs{75.10.Hk,75.30.Kz,75.70.-i,75.40.Mg}

\maketitle

\section{Introduction}
Artificial antiferromagnets, formed of nanostructured superlattices that are coupled antiferromagnetically,
have been the focus of many recent studies, due to their high potential for innovative technological
applications. These possible applications range from high-density recording technology \cite{Full03} to
spintronics devices \cite{Mang06} and magnetic refrigeration.\cite{Mukh09} 
Artificial antiferromagnets are heterostructures composed of ferromagnetic layers that are coupled
antiferromagnetically via spacers. This structure yields a high level of control over both the intra- 
and interlayer interactions, allowing for a tailoring of the physical properties.
Examples include [Co/Pt]/Ru, where ferromagnetic Co/Pt multilayers are periodically separated by Ru
layers,\cite{Hell03,Hell07} or Co/Cr, where thin Co films are separated by the spacer Cr.\cite{Mukh09} 
The magnetic moments of the ferromagnetic layers can thereby either be perpendicular (as it is
the case for Co/Pt with perpendicular anisotropy) or parallel (as encountered in Co films) to
the interface separating the ferromagnetic layers from the spacers.

[Co/Pt]/Ru with strong perpendicular anisotropy, as well as the related systems like [Co/Pt]/NiO, Co/Ir, and Fe/Au, have been called synthetic
metamagnets,\cite{Ross04a} as they can exhibit a regime with an antiferromagnetic phase at low external magnetic fields
and a paramagnetic phase at large fields. When changing the value of the magnetic field, plateaux show up
in the magnetic hysteresis of these films, due to the fact that different layers reverse their magnetization at different 
values of the field.\cite{Hell03} In fact, for an even number of identical ferromagnetic layers in a thin film exactly two metamagnetic transitions
are observed, as one of the outermost layers switches at a lower field than the internal layers.\cite{Ross04a} 
The same sequence of phases should also be realized for in-plane magnetization and strong anisotropy.\cite{Ross04b}

In a phenomenological approach it is customary to replace in systems with strong perpendicular anisotropy
the ferromagnetic multilayers by a single ferromagnetic layer.\cite{Ross04a,Ross04b} This naturally leads to a modeling of synthetic
metamagnets by an Ising metamagnet, i.e. a layered Ising model with ferromagnetic in-layer interactions
and antiferromagetic interactions between adjacent layers. The advantage of using a layered Ising model is that
this type of model is perfectly suited for a study of thermal properties through standard numerical methods.
However, modeling ferromagnetic multilayers by single ferromagnetic layers 
also has its restrictions as it does not allow a theoretical description
of the multidomain states, composed of metamagnetic stripe and bubble domains, that may form when applying a 
magnetic field to an artificial antiferromagnet with strong anisotropy.\cite{Hell07,Kise10}

Over the years Ising metamagnets have attracted quite some attention on their own.\cite{Harb73,Kinc75,Land81,Herr82,Land86,
Herr93,Hern93a,Hern93b,Selk95,Dasg95,Selk96,Pleim97,Gala98,Sant98,Zuko00,Sant00,More02,Gulp07,Geng08,deQu09,Lian10,Devi10}
Indeed, due to their simplicity, Ising metamagnets are ideal systems to study the properties of a tricritical point
that separates a discontinuous transition between the antiferromagnetic and paramagnetic phases at high fields
and low temperatures from a continuous transition between the same phases at low fields and high temperatures.
In addition, the mean-field prediction of a decomposition of the tricritical
point into a critical end point and a double critical end point \cite{Kinc75} has led to systematic numerical 
investigations of this possible scenario. Whereas many studies have concluded that this decomposition does not
take place in short-range models,\cite{Land86,Herr93,Sant98,Zuko00} the increase of the interlayer coordination, which brings the model closer 
to be of mean-field type, yields anomalies in the specific heat and the magnetization \cite{Selk95,Selk96} that resemble those
measured in FeBr$_2$.\cite{Azev95,Pell95} 

The general behavior of Ising metamagnets is by now well understood. However, all mentioned studies focused on bulk
systems and no systematic studies of metamagnetic properties of thin Ising films have been done. The recent 
interest in synthetic metamagnets warrants an in-depth understanding of metamagnets in confined
geometries, and we propose here a step in that direction by studying thin Ising metamagnetic films.

An additional motivation for our work comes from the recent analysis \cite{Mukh10} of 
non-equilibrium relaxation processes in Co/Cr superlattices. This study revealed that intriguing aging phenomena
take place in layered antiferromagnets. In order to better understand these observations a thorough study of
the dynamical properties of related theoretical models is needed. However, before being able to study in depth the
non-equilibrium properties of metamagnetic films, we found it necessary to first fully understand the equilibrium
properties of these systems. Therefore we focus in this paper on the equilibrium phase diagram of thin
metamagnetic films. The non-equilibrium properties of these systems will be discussed in a separate publication.

Our paper is organized in the following way. After having introduced  our model in Section II, we discuss
in Section III its equilibrium properties. Using the Wang-Landau scheme, we determine the density of
states (degeneracy) of our classical model and discuss its prominent features. This density of states is then
used for a canonical analysis of the system where the focus is on the two phase transitions that are observed
in thin film geometry when increasing the external magnetic field. As the Wang-Landau scheme is restricted to small
systems, we supplement our study by standard Monte Carlo simulations of larger systems. 
From these numerical data we derive the phase diagram of thin metamagnetic films and show that in thin films
a new phase transition line shows up that separates two ordered phases. This transition, which is
absent in bulk systems, is found to be 
discontinuous and to end in a critical end point.
Finally, in Section IV we discuss our results and conclude.
 
\section{Model and methods}
We consider a layered lattice model on a cubic lattice where every lattice point $i$ is characterized by an Ising
spin, $S_i = \pm 1$. The interactions between nearest neighbor spins are ferromagnetic in the planes perpendicular
to the $z$ axis. These two-dimensional planes are coupled antiferromagnetically in the $z$ direction. Adding an
external magnetic field of strength $B$, the Hamiltonian of our system is then given by
\begin{equation} \label{eq:hamiltonian}
{\mathcal H} = - J_{xy} \sum\limits_{\langle i,j\rangle} S_i S_j + J_z \sum\limits_{(i,j)} S_i S_j - B \sum\limits_i S_i = E - B M~,
\end{equation}
where we have introduced the internal energy $E = - J_{xy} \sum\limits_{\langle i,j\rangle } S_i S_j + J_z \sum\limits_{(i,j)} S_i S_j$ 
and the magnetization $M = \sum\limits_i S_i$.
The sums over $\langle i,j\rangle$ resp. $(i,j)$ are sums over nearest neighbor pairs in the plane resp. along the $z$ direction. 
The intralayer and interlayer coupling strengths are given by $J_{xy} > 0$ and $J_z > 0$.

In contrast to previous studies we focus on thin films which are realized by imposing free boundary conditions in the
$z$ direction, whereas in the $x$ and $y$ directions we have periodic boundary conditions. In accordance with the
synthetic metamagnets we consider rather few layers, typically $L = 8$ or $L=10$ (we restrict ourselves to even numbers
of layers). For the case of periodic boundary conditions in all three directions, which has been studied extensively
in the past, the system undergoes a metamagnetic transition between an antiferromagnetic phase and a paramagnetic phase
when increasing the field strength. This transition is discontinuous for high fields and low temperatures and continuous 
for low fields and high temperatures. It is expected that an additional transition shows up when considering films,
this transition being characterized by the alignment of the magnetization of one of the outermost planes 
with the external field.\cite{Hell03,Ross04a} 

We study this system using different simulation techniques. In order to elucidate its static properties we compute the
degeneracy (or density of states) $\Omega(E,M)$ as a function of internal energy $E$ and magnetization $M$ using the Wang-Landau
scheme.\cite{Wang01a,Wang01b} The degeneracy can then be used for a standard canonical analysis, with the partition 
sum as a function of temperature $T$ and magnetic field $B$ being given by (we choose units for which $k_B = 1$)
\begin{equation} \label{eq:Z}
Z(T,B) = \sum\limits_E \sum\limits_M \Omega(E,M) e^{- E/T + B M/T}~.
\end{equation}
All global quantities of interest (like the mean magnetization, the mean energy, the susceptibility, 
and the specific heat) then follow
from derivatives of $Z$ with respect to $T$ and $B$. It is well known that the Wang-Landau scheme usually yields a very good
estimate for the degeneracy, allowing for a detailed canonical analysis. However, only rather small systems can be studied
in this way. We therefore supplement our study with traditional importance sampling simulations of larger system sizes, using
the Metropolis algorithm. We thereby study thin films with layers that contain $L \times L$ spins, with $L = 8$, 16, 32, and 64.
Relevant quantities are computed for a large number of temperatures and field strengths; in order to completely cover the
phase diagram, the increment between successive $T$ and $B$ values is typically 0.01.

\section{Equilibrium properties}

We focus in the following on the thermal equilibrium properties of thin Ising ferromagnets.
The statistical physics treatment of our classical spin
models allows for an in-depth discussion of its properties as a function of all relevant parameters, which in our case
are the temperature and the strength of the magnetic field.

\subsection{The density of states}

As already mentioned we compute for our smaller systems the density of states $\Omega(E,M)$, i.e. the number of microscopic 
configurations that have the same internal energy $E$ and the same magnetization $M$. Having this two-dimensional histogram
at our disposal, we can then compute all relevant global thermodynamic quantities from the partition sum (\ref{eq:Z}) and its derivatives
by simply inserting the numerical values for $T$ and $B$.

\begin{figure}[h]
\includegraphics[angle=0,width=6.3in]{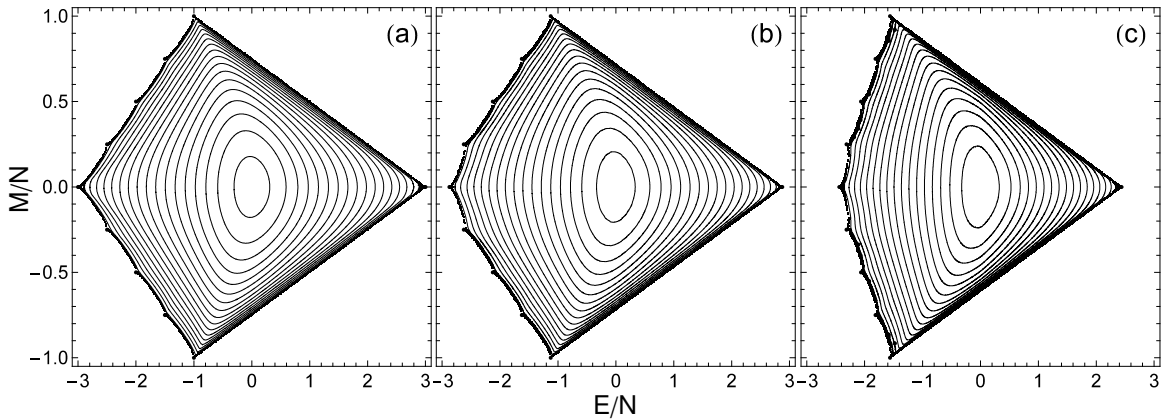}
\caption{Microcanonical entropy $S(e,m)$ as a function of energy density $e = E/N$ and magnetization density $m = M/N$ for systems
containing $N = 8 \times 8 \times 8$ spins. (a) Periodic boundary conditions in all three directions and $J_{xy} = J_z$,
(b) free surfaces in $z$ direction with $J_{xy} = J_z$, and (c) free surfaces in $z$ direction with $J_{xy} = 2 J_z$.
The increment between two successive contour lines is 20. The supports of the entropy surfaces are indicated by the larger points.
}\label{fig1}
\end{figure}

Before doing that we briefly discuss the density of states itself. Fig. \ref{fig1} shows the microcanonical entropy
$S(E,M) =  \ln \Omega(E,M)$ as a function of the energy density $e = E/N$ and the magnetization density $m = M/N$
(here $N$ is the total number of spins in our systems) for three different cases. Whereas in (a) we consider a system 
with periodic boundary conditions composed
of $N = 8 \times 8 \times 8$ spins, with $J_{xy} = J_z$, in (b) and (c) we show two systems of $N = 8 \times 8 \times 8$ sites
with free boundary condition in $z$ direction, the different systems having different relative strengths 
of the interactions: $J_{xy} = J_z$ in (b) and 
$J_{xy} = 2 J_z$ in (c).

In a microcanonical analysis one infers the physical property of a system from a direct study of the microcanonical entropy $S(e,m)$.
\cite{Gros01,Plei05} Most investigations of this type focused on spin models with ferromagnetic nearest neighbor interactions,
as for example the standard Ising or Potts models,\cite{Kast00,Gros00,Huel02,Plei04,Hove04,Behr05,Rich05,Behr06,Kast09}
or on polymer models.\cite{Jung06,Moed10}
Due to its complicated interactions, the microcanonical entropy of the Ising metamagnet is more complex as, for example, that of the
standard nearest neighbor Ising model,\cite{Plei05} see Fig. \ref{fig1}. Still some of its features and properties are readily understood.

Firstly, there are obvious properties that are independent of the boundary conditions and of the interaction ratio
$J_z/J_{xy}$. The ground state is two-fold degenerate and has magnetization zero (recall that we only study systems with even numbers of layers),
with the fully ordered layers pointing alternatively in up or down direction. Obviously, a given ground state can
be changed into the other ground state by multiplying all the spins by $-1$. This symmetry of the internal energy also
shows up in the symmetry of the entropy surfaces with respect to the $M=0$ line.
Interestingly, the entropy surfaces all display prominent kinks at magnetizations $M/N = \pm 2\mu/L$, with $\mu=0, \cdots, L/2$, corresponding
to configurations with fully ordered planes where $\mu$ planes have been flipped with respect to the ground state.

Changing the interaction ratio $J_z/J_{xy}$, see Fig.\ \ref{fig1}b and \ref{fig1}c, mainly changes the range of accessible
energies. As a result, smaller values of $J_z/J_{xy}$ yield a compressed, but otherwise unchanged, entropy surface.

At a first look, changing the boundary condition in $z$ direction also seems to only have minor effects on $s(e,m)$. 
However, one notes that the energy difference
between the ground state and the state with one flipped layer, with $M/N= \pm 2/L = \pm 1/4$ in Fig. \ref{fig1}, is smaller for
the free boundaries in $z$ direction than for the periodic boundaries, compare Fig.\ \ref{fig1}a and \ref{fig1}b. As we already argued
in the introduction (and as we will see later in the canonical analysis), our metamagnetic films should display two metamagnetic
transitions, a first transition at low magnetic fields, where only one surface layer is flipped, and a second transition
at higher fields where the remaining layers pointing opposite to the magnetic field are flipped. In contrast, a system with periodic
boundary conditions in all three directions is known to undergo a single metamagnetic transition.\cite{Kinc75}
With this in mind, it seems surprising
that very similarly looking entropy surfaces should yield these different phase transition sequences.

\begin{figure}[h]
\includegraphics[angle=270,width=5.6in]{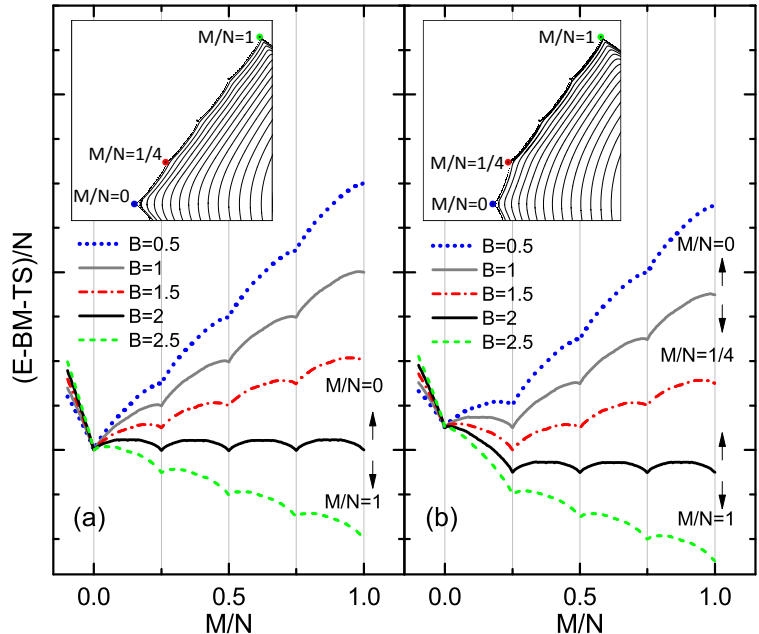}
\caption{(Color online) "Microcanonical" quantity $(E - B M - T S)/N$ as a function of $M/N$.
The canonical correspondence of this quantity is the Helmholtz free energy density. The global minimum of the
plotted quantity allows to understand the sequence of phases that show up when changing
the strength of the magnetic field. The systems contain $N = 8 \times 8 \times 8$ spins, with (a) periodic
boundary conditions and (b) free boundary conditions in $z$ direction. The insets show the relevant parts
of the microcanonical entropy, see main text.
}\label{fig2}
\end{figure}

In fact, it is the decrease of the energy difference between the ground state and the 
states with fully ordered planes and magnetization $M/N= \pm 2/L$ that ultimately is responsible for the emergence
of this additional transition. We show that in Fig. \ref{fig2} where we plot 
for systems with $J_z = J_{xy}$ the quantity $(E - B M - T S)/N$ as a function
of $M/N$, with $T=1$ and different values of $B$. In the canonical ensemble the corresponding quantity is of course
the Helmholtz free energy density which is minimal for the stable state. But even without making a canonical
average one can read off the stable phase from the "microcanonical" quantity plotted in Fig. \ref{fig2}. For the system with periodic boundary
conditions, see Fig. \ref{fig2}a, our quantity is minimal for $M=0$ and low fields. When the field strength exceeds $B=2$,
a metamagnetic phase transition takes place such that the stable phase is now the paramagnetic phase with $M/N = 1$.
Conversely, for the thin film geometry, see Fig. \ref{fig2}b, a first transition to a 
phase with one flipped layer and magnetization $M/N= 2/L = 1/4$ shows up at $B=1$, as can be seen when studying the global minimum
of $E - B M - T S$, followed by a second discontinuous transition at $B=2$. In this way, the sequence of
phase transitions can indeed be unraveled in a microcanonical analysis.

\subsection{Thermal quantities}
Even so the microcanonical analysis allows to determine the sequence of phases, an in-depth study
of the thermal properties of the metamagnetic films warrants a canonical analysis of the standard
quantities as a function of temperature and magnetic field. In the following, we discuss small systems
for which the density of states can be obtained numerically, as discussed in the previous subsection.
This density of states $\Omega(E,M) = e^{S(E,M)}$ is then used for the computation of the partition sum and related quantities.
Thus for a quantity $Q(E,M)$ the thermal average at temperature $T$ and field strength $B$ is 
\begin{equation}
\langle Q \rangle = \sum\limits_E \sum\limits_M Q(E,M) \Omega(E,M) e^{-(E - B M)/T}/Z(T,B)~,
\end{equation}
where the partition function $Z(T,B)$ is given by Eq. (\ref{eq:Z}). In order to elucidate the
thermal properties of thin metamagnets we studied the average magnetization
$\langle M \rangle$, the average energy $\langle E \rangle$, the magnetic susceptibility $\chi = \frac{1}{N T} \left[ \langle M^2 \rangle - \langle M 
\rangle^2 \right]$, and the specific heat $C = \frac{1}{N T^2} \left[ \langle {\mathcal H}^2 \rangle - \langle {\mathcal H} \rangle^2 \right]$.

We supplement this canonical analysis of small systems based on the density of states by standard canonical
simulations of larger systems, with the same number of layers but different number of spins in the layers.
This allows us to assess the finite-size effects and to extrapolate to films with a large number of spins in every layer.
Standard importance sampling Monte Carlo simulations often yield results of lesser quality than the canonical analysis based
on the Wang/Landau scheme. On the other hand larger systems can easily be simulated. In addition, we also can look at additional
quantities which are not accessible when starting from the degeneracy as a function of total magnetization and total energy.
Thus we will also study the layer susceptibility
\begin{equation} \label{eq:surf_chi}
\chi(z) = \frac{1}{L^2 T} \left[ \langle M^2(z) \rangle - \langle M(z) \rangle^2 \right]
\end{equation}
where $M(z)$ is the magnetization of layer $z$. Of special interest is of course the susceptibility $\chi(1)$
of the layer $z=1$ that flips in the direction of the magnetic field when crossing the low field phase transition line.
Similarly, we also study the layer specific heat
\begin{equation} \label{eq:surf_C}
C(z) = \frac{1}{L^2 T^2} \left[ \langle {\mathcal H}^2(z) \rangle - \langle {\mathcal H}(z) \rangle^2 \right]
\end{equation}
where ${\mathcal H}(z) = - J_{xy} \sum\limits_{\langle i,j\rangle_z } S_i S_j - B \sum\limits_{i}^{layer ~z} S_i$ is the in-layer contribution
to the energy in layer $z$.

\begin{figure}[h]
\includegraphics[angle=0,width=4.5in]{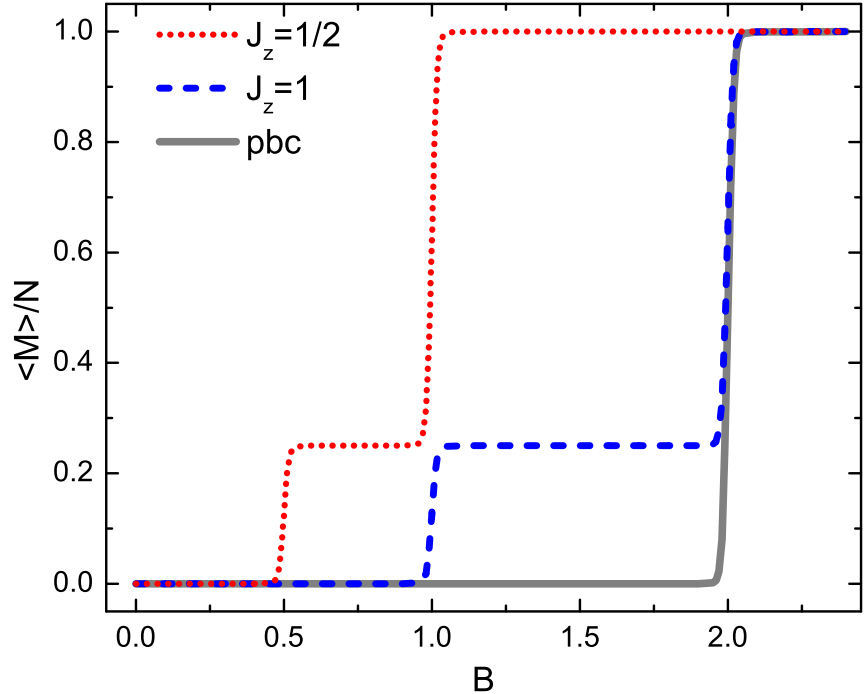}
\caption{(Color online) Average magnetization as a function of magnetic field strength $B$ at temperature $T = 1$. The magnetization
is shown for thin metamagnetic films with different interaction ratios as well as for a sample with periodic boundary conditions (pbc)
and $J_z = 1$. The three-dimensional samples contain $N = 8^3$ spins, with the interlayer interaction having the strength
$J_{xy}=1$.
}\label{fig3}
\end{figure}

Fig. \ref{fig3} highlights the expected difference in the field dependence of the magnetization between thin
films and bulk systems (i.e. systems with periodic boundary conditions). For thin films a first
transition takes place for $B \approx J_z$, where one of the outer layers completely flips, followed by a second transition
at $B \approx 2 J_z$, where the remaining layers align with the magnetic field. At $T=0$ these transitions take place exactly at 
$B = J_z$ and $B = 2 J_z$. For larger $T$ than that used in Fig. \ref{fig3}, the total magnetization after the flipping of the
outer layer is slightly lower than $N/4$, due to thermal fluctuations.

\begin{figure}[h]
\includegraphics[angle=270,width=6.3in]{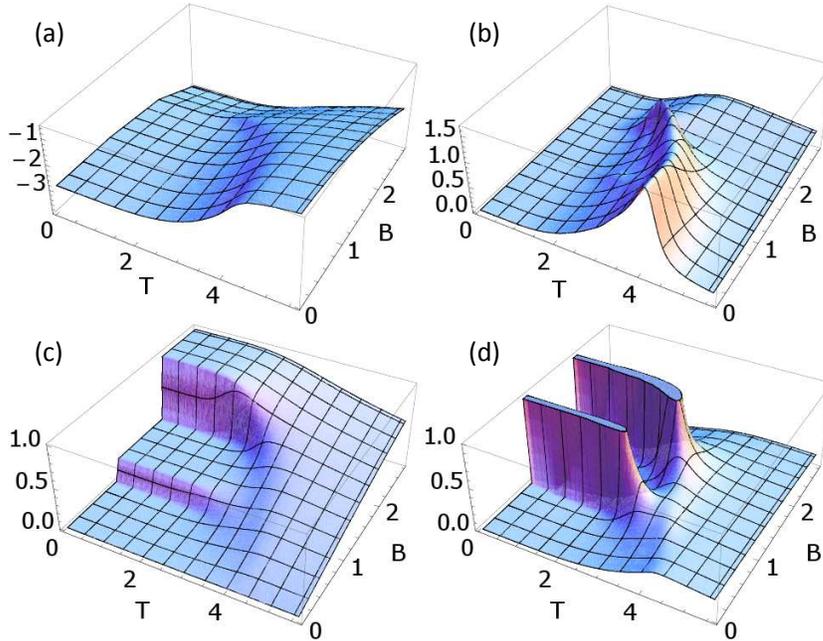}
\caption{(Color online) (a) Energy density, (b) specific heat, (c) magnetization density, and (d) susceptibility as a function
of $T$ and $B$ for a thin film composed of $N = 8^3$ spins, with $J_z = J_{xy} = 1$.
}\label{fig4}
\end{figure}

We discuss the $T$ and $B$ dependence of our thermal quantities in Fig. \ref{fig4} for a thin film of
eight layers with $J_z = J_{xy} = 1$. The two phase transitions are readily seen
in the changes of the magnetization density (plateaux in Fig. \ref{fig4}c) and the susceptibility (lines of maxima in Fig. \ref{fig4}d). 
One notes that for small $T$ the change in the magnetization is very abrupt, pointing to a discontinuous
transition, whereas for larger $T$ this change is much more gradual, in agreement with a continuous
transition. This change of the nature of the transition also shows up in the susceptibility. Two lines of
maxima, one being hardly visible on the scale of the figure, are also observed in the specific heat, see Fig. \ref{fig4}b.

\begin{figure}[h]
\includegraphics[angle=0,width=4.5in]{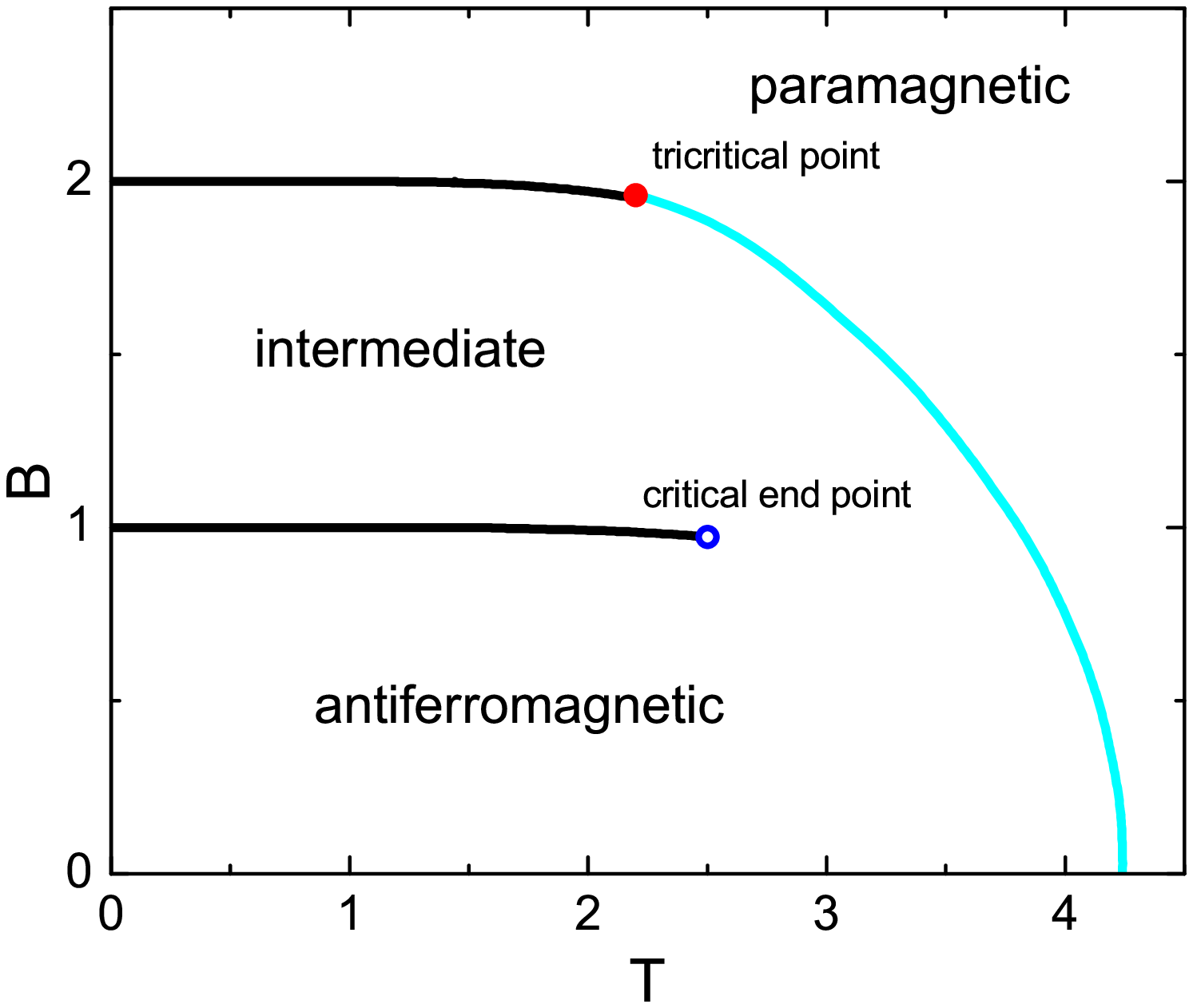}
\caption{(Color online)  Phase diagram for the thin Ising metamagnet film with eight layers. The different
transition lines result from the maxima of the (global and layer) susceptibility and specific heat. Black lines indicate
a discontinuous transition, whereas the continuous transition is shown as a grey (cyan online) line. We observe a tricritical point
(filled point) as well as a critical end point (open point).
}\label{fig5}
\end{figure}

\begin{figure}[h]
\includegraphics[angle=0,width=5.3in]{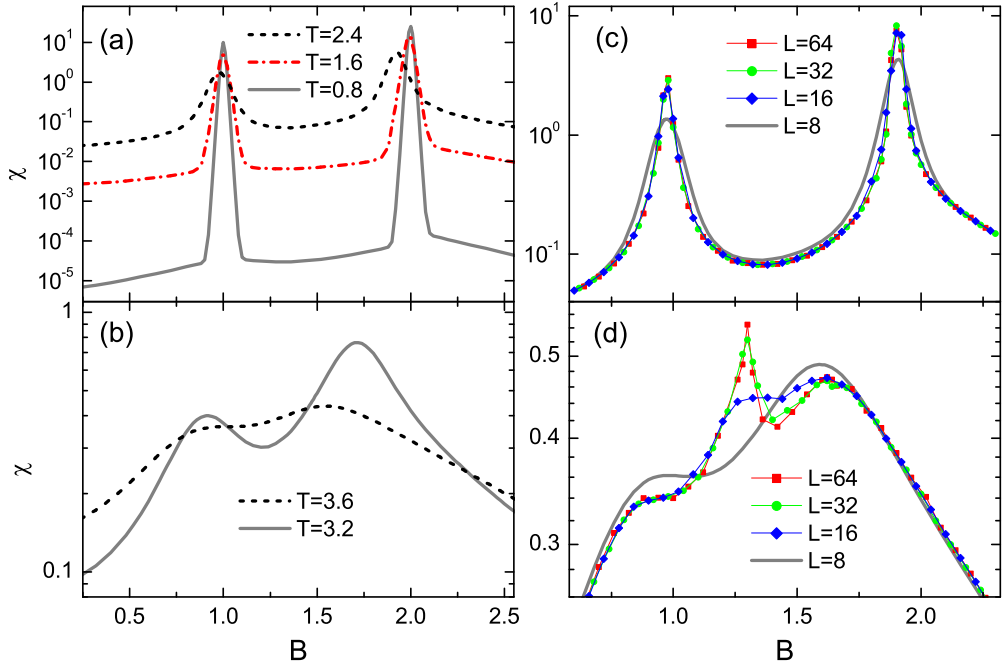}
\caption{(Color online) 
Susceptibility of thin films with 8 layers as a function of magnetic field strength for various temperatures. 
(a,b) Data derived from the density of states of a small system with $8\times8$ spins per layer. At low temperatures
the phase transition is discontinuous, as revealed by the characteristic form of the peaks.
Around $T \approx 2.2$ the form of the peaks changes, indicating
that the transition has become continuous. 
(c) Susceptibility for $T=2.5$ obtained
for systems with different numbers of spins in the layers.
Whereas the height of the peaks changes with system size, the location of the peaks does not show a dependence on the sizes of the
layers. Symbols indicate data obtained using the Metropolis algorithm. (d) The same as in (c), but now for $T=3.5$. For larger systems
a critical peak, which indicates the transition to the paramagnetic phase, emerges that is not visible for the smallest systems. 
The peak at $B \approx 1.6$ is a non-critical peak. 
Error bars are comparable to the sizes of the symbols.
}\label{fig6}
\end{figure}

\begin{figure}[h]
\includegraphics[angle=0,width=4.5in]{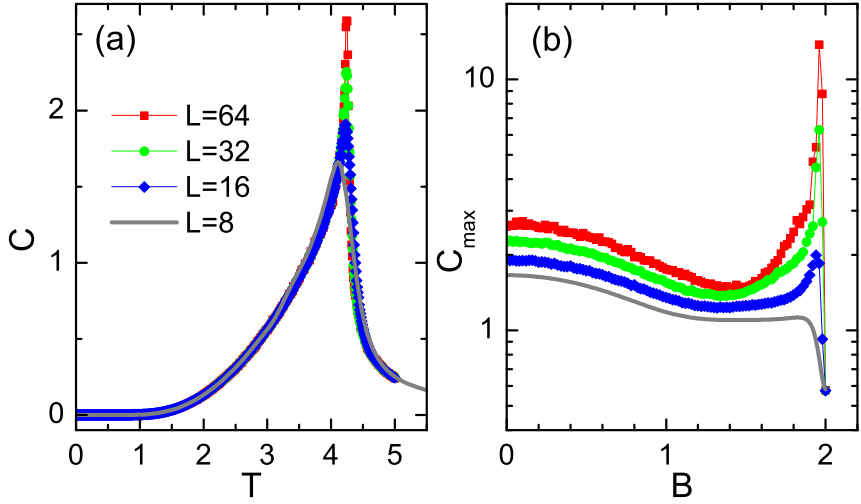}
\caption{(Color online) (a) Specific heat of an eight-layer system as a function of $T$ for $B = 0$ and various sizes of the layers.
The peaks indicate the transition between the ordered and paramagnetic phases.
(b) The maximum height of the specific heat for a fixed value of the magnetic field strength $B$. This maximum value
is achieved at the transition to the paramagnetic phase. The strong peak around $B \approx 1.96$ reveals the change of
the order of the transition when crossing the tricritical point.
}\label{fig7}
\end{figure}

\begin{figure}[h]
\includegraphics[angle=0,width=5.3in]{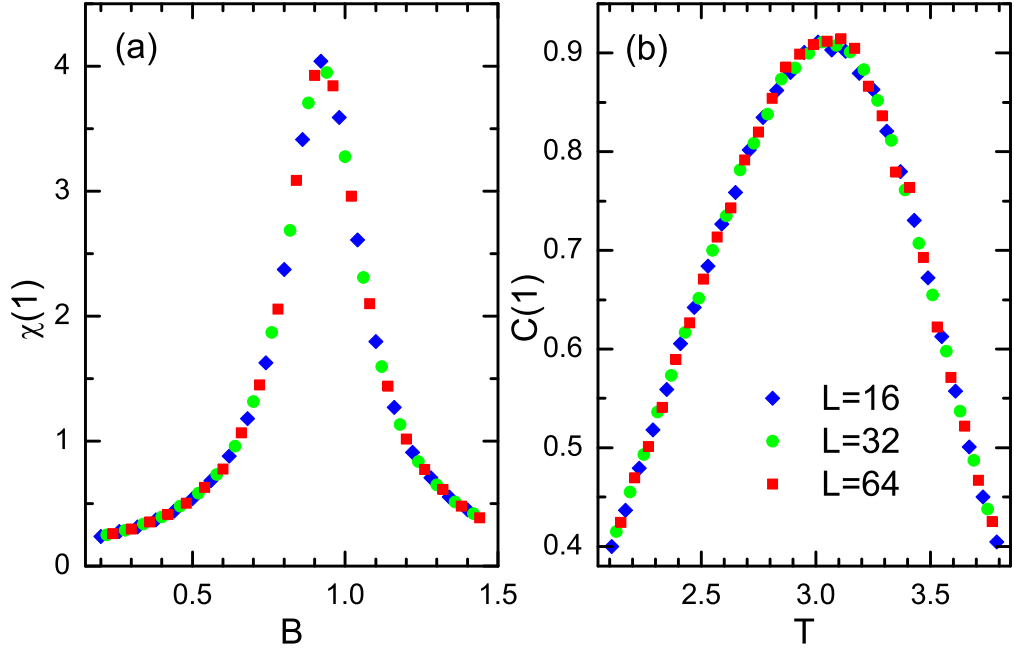}
\caption{(Color online) (a) Surface susceptibility (\ref{eq:surf_chi}) at $T=3$ and (b) surface specific heat (\ref{eq:surf_C}) at $B = 0.5$ 
for the surface layer that aligns with the magnetic field. Data obtained for different system sizes are shown.
These maxima are non-critical and size independent, indicating that the reversal of the surface magnetization is gradual for these
values of the system parameters.
}\label{fig8}
\end{figure}

Based on the positions of the maxima in the response functions, see Figures \ref{fig6}, \ref{fig7},
and \ref{fig8}, we obtain the phase diagram 
for a thin metamagnetic film composed of 8 layers shown in Fig. \ref{fig5}.
We first note the expected presence of three different phases: the paramagnetic phase at high temperatures and high fields,
the antiferromagnetic phase at low temperatures and low fields, and a phase which has a non-zero magnetization at intermediate fields
and low temperatures. Similar to what is observed in the phase diagram of the bulk system, the transition between the ordered
and paramagnetic phases is discontinuous at low temperatures and continuous at high temperatures, with a tricritical point
separating these two regimes. We locate this point at $T=2.20(1)$  and $B=1.96(1)$. 
At low temperatures the transition between the
antiferromagnetic phase and the intermediate phase is also discontinuous. This transition, however, does not extend 
all the way up to the phase transition line separating the ordered phase from the paramagnetic phase, but instead ends at
a critical end point located at $T=2.50(5)$ and $B=0.98(1)$, see below.

Increasing the thickness of the film leaves the phase diagram qualitatively unchanged. A slight shift in the phase transition lines,
especially at high temperatures and low fields, is observed when changing the number of layers in the film. This shift of the
critical temperature of a film as a function of
thickness is of course expected and has been studied extensively, both theoretically and
experimentally, in the absence of external magnetic fields (see, for example, [\onlinecite{Barb83,Li92,Schi96,Zhan01,Plei04a}]).

In Figures \ref{fig6}, \ref{fig7}, and \ref{fig8} we have a closer look at the different response functions
used for the construction of the phase diagram. 
At low temperatures,
see  Fig. \ref{fig6}a, the peaks in the susceptibility as a function of the magnetic field strength are very pronounced and sharp,
indicating the discontinuous character of the transitions. The form of the peaks change around $T=2.2$, as here the order of
the transition changes from discontinuous to continuous. Above $T = 2.2$, see Fig. \ref{fig6}c, the height of the peaks shows the
expected size dependence of a continuous transition. Fig. \ref{fig6}d shows the total susceptibility at the rather 
high temperature
of $T=3.5$. Increasing the system size reveals the emergence of a critical peak at $B \approx 1.3$. This peak,
which coincides which the transition to the paramagnetic phase, only shows
up as a shoulder in the smaller systems. The additional peak at $B \approx 1.6$ is a non-critical one and is similar
to that observed in the bulk system.

Another way to monitor the phase transition lines is through a study of the specific heat as a function of
temperature. As an example Fig. \ref{fig7}a shows the specific heat for vanishing magnetic field. Only one
critical peak is observed that results from the phase transition between the ordered low temperature phase
and the disordered high temperature phase. In Fig. \ref{fig7}b we show the field dependence of the maximum of
the specific heat along the phase transition line to the paramagnetic phase. For low fields this transition
is continuous and the specific heat height shows the expected finite size scaling of an ordinary critical
point. For a fixed system size the specific heat exhibits a strong peak around $B \approx 1.96$ which is
due to the change of the order of the phase transition when crossing the tricritical point.

The standard way to locate the tricritical point is to monitor the hysteris observed when crossing the
phase transition line at low fields \cite{Herr93,Zuko00}, as the hysteresis loop vanishes when approaching
the tricritical point. We found it useful to monitor in addition the location of a non-critical high field
peak observed for temperatures
larger than the temperature of the tricritical point. This non-critical peak is also observed
in metamagnetic bulk systems, both in simulations \cite{Pleim97} and in experiments \cite{Bin00}. The merging of this non-critical
line with the phase transition line separating the paramagnetic phase
from the ordered phases allows to reliably estimate the location of the tricritical point. Another estimate
can be obtained by monitoring the strong increase of the
peak heights of response functions when approaching the tricritical point, as shown in Fig. \ref{fig7}b.
Taking all this into account, we estimate the tricritical point of our thin film to be at $T=2.20(1)$ and $B=1.96(1)$.

An interesting property of the phase diagram of thin metamagnetic films emerges when studying
response functions at high temperatures and low fields. 
Whereas for low temperatures the response functions display a characteristic peak 
with a height that depends on the system size, for higher temperatures 
the response functions do no longer display any dependence on the size of the system.
As shown in Fig.\ \ref{fig8} for the surface response functions, 
system size dependent peaks are no longer encountered in the ordered phase for temperatures $T \geq 2.50$ and fields
$B \geq 0.98$. This excludes the existence of a line of continuous phase transitions at high temperatures and
low fields, indicating that the line of discontinuous transitions ends in a critical end point.
Therefore, instead of crossing a phase transition line, with its characteristic singularities, one can
go from the antiferromagnetic phase at low temperatures and fields to the intermediate
phase with one flipped layer in a smooth and gradual way. This is in complete analogy to the well known
behavior of water where one can go in a similar way from vapor to liquid without a phase transition.
Based on our data we locate this critical end point at $T=2.50(5)$ and $B=0.98(1)$.

%
%
%
%

\section{Discussion and conclusion}
The numerous possible applications have recently yielded a strong interest in artificial antiferromagnets and synthetic metamagnets.
Remarkably, the layered structure of these materials allows for a high level of control of their physical properties through the fine-tuning
of the strengths of both the inter- and intralayer interactions.

In order to get a better understanding of some of these properties we presented a study of the equilibrium properties
of thin Ising metamagnetic films. Indeed, neglecting the internal structure of ferromagnetic multilayers, one can model 
artificial antiferromagnets with strong perpendicular anisotropy by a layered Ising antiferromagnet. If one notes in
addition that the synthetic metamagnets are usually composed of only a few repetition of the superlattice structure, one
is then naturally led to consider Ising metamagnets in thin film geometry.

Our work allowed us to study in detail the phase diagram of these systems as a function of temperature and of the strength
of the applied external magnetic field. Earlier studies of related phenomenological models revealed \cite{Ross04a,Ross04b} 
that for an even number of layers a layered antiferromagnet
has three different $T=0$ states, depending on the strength of the magnetic field. The antiferromagnetic structure, stable at
low fields, goes over into a different ordered structure at intermediate field strengths. 
In this intermediate phase one of the surface layers aligns with the direction of
the magnetic field, the other layers remaining unchanged. At larger fields the intermediate phase is replaced by the
paramagnetic phase where all layers are parallel to the magnetic field. This intermediate phase is absent in the
corresponding bulk system and is therefore characteristic for thin metamagnetic films. 

Using the Wang-Landau algorithm and importance sampling Monte Carlo simulations we thoroughly studied the equilibrium 
properties of our system. The Wang-Landau method yields the microcanonical entropy (or, alternatively, the degeneracy)
as a function of magnetization and internal energy. The microcanonical entropy surface is rather complicated and reveals
interesting features that we discussed in some detail. Thus we were able to relate the appearance of the intermediate
phase to a subtle change in the microcanonical entropy when changing the boundary condition.

Our canonical analysis of different thermal quantities, as for example the global and layer magnetization,
the global and layer susceptibility as well as the global and layer specific heat, allows us to determine the
temperature $-$ magnetic field phase diagram shown in Fig.\ \ref{fig5}, which is the main result of our study.
Interestingly, the discontinuous phase transition between the antiferromagnetic and the intermediate phase ends in
a critical end point. It is therefore possible to pass from one phase to the other without undergoing a phase
transition. Experimental studies on systems with strong perpendicular anisotropy should be able to verify this
intriguing feature of thin Ising metamagnets.

As already mentioned in the Introduction one of the motivations for our study was the recent investigation \cite{Mukh10}
of non-equilibrium relaxation and aging phenomena in Co/Cr superlattices. This study revealed intriguing non-equilibrium
features that warrant a better understanding. As a first step in that direction we started a study of the
non-equilibrium properties of thin Ising metamagnets, but soon realized that a more complete understanding of
the equilibrium properties of these systems is needed before being able to develop better insights
into the more complex situation of relaxation
far from equilibrium. With the new knowledge of the equilibrium properties reported in this paper we are now well prepared to
better understand these complicated non-equilibrium processes. The results of this investigation will be the
subject of a forthcoming paper.

\begin{acknowledgments}
We thank Christian Binek and Tatha Mukherjee for insightful discussions.
This work was supported by the US National
Science Foundation through DMR-0904999.
M.P. thanks the Max-Planck-Institut f\"{u}r Physik komplexer Systeme
in Dresden, Germany, for the hospitality
during the completion of this work.
\end{acknowledgments}


\begin{thebibliography}{99}

\bibitem{Full03} 
E. E. Fullerton, D. T. Margulies, N. Supper, Do Hoa, M. Schabes, A. Berger, and A. Moser,    
IEEE Trans. Magn. {\bf 39}, 639 (2003).

\bibitem{Mang06} S. Mangin, D. Ravelosona, J. A. Katine, M. J. Carey, B. D. Terris, and E. E. Fullerton,
Nature Mater. {\bf 5}, 210 (2006).

\bibitem{Mukh09} T. Mukherjee, S. Sahoo, R. Skomski, D. J. Sellmyer, and Ch. Binek, 
Phys. Rev. B {\bf 79}, 144406 (2009).

\bibitem{Hell03} O. Hellwig, T. L. Kirk, J. B. Kortright, A. Berger, and E. E. Fullerton,
Nature Mater. {\bf 2}, 112 (2003).

\bibitem{Hell07} O. Hellwig, A. Berger, J. B. Kortright, and E. E. Fullerton,
J. Magn. Magn. Mater. {\bf 319}, 13 (2007).

\bibitem{Ross04a} U. K. R\"{o}{\ss}ler and A. N. Bogdanov,
J. Magn. Magn. Mater. {\bf 269}, L287 (2004).

\bibitem{Ross04b} U. K. R\"{o}{\ss}ler and A. N. Bogdanov,
Phys. Rev. B {\bf 69}, 094405 (2004).

\bibitem{Kise10} N. S. Kiselev, C. Bran, U. Wolff, L. Schultz, A. N. Bogdanov,
O. Hellwig, V. Neu, and U. K. R\"{o}{\ss}ler, Phys. Rev. B {\bf 81}, 054409 (2010).

\bibitem{Harb73} F. Harbus and H. E. Stanley, Phys. Rev. B {\bf 8}, 1141 (1973).

\bibitem{Kinc75} J. M. Kincaid and E. G. D. Cohen, 
Phys. Rep. {\bf 22}, 57 (1975).

\bibitem{Land81} D. P. Landau and R. H. Swendsen, Phys. Rev. Lett. {\bf 46}, 1437 (1981).

\bibitem{Herr82} H. J. Herrmann, E. B. Rasmussen, and D. P. Landau,
J. Appl. Phys. {\bf 53}, 7994 (1982).

\bibitem{Land86} D. P. Landau and R. H. Swendsen, Phys. Rev. B {\bf 33}, 7700 (1986).

\bibitem{Herr93} H. J. Herrmann and D. P. Landau, Phys. Rev. B {\bf 48}, 239 (1993).

\bibitem{Hern93a} L. Hern\'{a}ndez, H. T. Diep, and D. Bertrand, Europhys. Lett. {\bf 21}, 711 (1993).

\bibitem{Hern93b} L. Hern\'{a}ndez, H. T. Diep, and D. Bertrand, Phys. Rev. B {\bf 47}, 2602 (1993).

\bibitem{Selk95} W. Selke and S. Dasgupta, J. Magn. Magn. Mater. {\bf 147}, L245 (1995).

\bibitem{Dasg95} S. Dasgupta, J. Stat. Phys. {\bf 81}, 837 (1995).

\bibitem{Selk96} W. Selke, Z. Phys. B {\bf 101}, 145 (1996).

\bibitem{Pleim97} M. Pleimling and W. Selke, Phys. Rev. B {\bf 56}, 8855 (1997).

\bibitem{Gala98} S. Galam, C, S. O. Yokoi, and S. R. Salinas, 
Phys. Rev. B {\bf 57}, 8370 (1998).

\bibitem{Sant98} M. Santos and W. Figueiredo, Phys. Rev. B {\bf 58}, 9321 (1998).

\bibitem{Zuko00} M. \v{Z}ukovi\v{c} and T. Idogaki, Phys. Rev. B {\bf 61}, 50 (2000).

\bibitem{Sant00} M. Santos and W. Figueiredo, Phys. Rev. E {\bf 62}, 1799 (2000).

\bibitem{More02} A. F. S. Moreira, W. Figueiredo, and V. B. Henriques, 
Eur. Phys. J. B {\bf 27}, 153 (2002).

\bibitem{Gulp07} G. Gulpinar, D. Demirhan, and F. Buyukkilic, Physica A {\bf 383}, 372 (2007).

\bibitem{Geng08} J. Geng, G. Wei, and H. Miao, J. Magn. Magn. Mater {\bf 320}, 1010 (2008).

\bibitem{deQu09} S. L. A. de Queiroz, Phys. Rev. E {\bf 80}, 041125 (2009)

\bibitem{Lian10} Y.-Q. Liang, G.-Z. Wei, X.-J. Xu, and G.-L. Song,
J. Magn. Magn. Mater {\bf 322}, 2219 (2010).

\bibitem{Devi10} B. Deviren and M. Keskin, Phys. Lett. A {\bf 374}, 3119 (2010).

\bibitem{Azev95} M. M. P. de Azevedo, Ch. Binek, J. Kushauer, W. Kleemann, and
D. Bertrand, J. Magn. Magn. Mater. {\bf 140-144}, 1557 (1995).

\bibitem{Pell95} J. Pelloth, R. A. Brand, S. Takele, M. M. Pereira de Azevedo, W. Kleemann,
Ch. Binek, J. Kushauer, and D. Bertrand, Phys. Rev. B {\bf 52}, 15372 (1995).

\bibitem{Mukh10} T. Mukherjee, M. Pleimling, and Ch. Binek, Phys. Rev. B {\bf 82}, 134425 (2010).


\bibitem{Wang01a} F. Wang and D. P. Landau, Phys. Rev. Lett. {\bf 86}, 2050 (2001).

\bibitem{Wang01b}  F. Wang and D. P. Landau, Phys. Rev. E {\bf 64}, 056101 (2001).

\bibitem{Gros01} D. H. E. Gross,
{\em Microcanonical Thermodynamics: Phase Transitions in ’Small’ Systems}, Lecture
Notes in Physic 66 (World Scientific, 2001).

\bibitem{Plei05} M. Pleimling and H. Behringer, Phase Transitions {\bf 78}, 787 (2005).

\bibitem{Kast00} M. Kastner, M. Promberger, and A. H\"{u}ller, J. Stat. Phys. {\bf 99}, 1251 (2000).

\bibitem{Gros00} D. H. E. Gross and E. V. Votyakov, Eur. Phys. J. B {\bf 15}, 115 (2000).

\bibitem{Huel02} A. H\"{u}ller and M. Pleimling, Int. J. Mod. Phys. C {\bf 13}, 947 (2002).

\bibitem{Plei04} M. Pleimling, H. Behringer, and A. H\"{u}ller, Phys. Lett. A {\bf 328}, 432 (2004).

\bibitem{Hove04} J. Hove, Phys. Rev. E {\bf 70}, 056707 (2004).

\bibitem{Behr05} H. Behringer, M. Pleimling, and A. H\"{u}ller, J. Phys. A: Math. Gen. {\bf 38}, 973 (2005).

\bibitem{Rich05} A. Richter, M. Pleimling, and A. H\"{u}ller, Phys. Rev. E {\bf 71}, 056705 (2005).

\bibitem{Behr06} H. Behringer and M. Pleimling, Phys. Rev. E {\bf 74}, 011108 (2006).

\bibitem{Kast09} M. Kastner and M. Pleimling, Phys. Rev. Lett. {\bf 102}, 240604 (2009).

\bibitem{Jung06} C. Junghans, M. Bachmann, and W. Janke, Phys. Rev. Lett. {\bf 97}, 218103 (2006). 

\bibitem{Moed10} M. M\"{o}ddel, W. Janke, and M. Bachmann, Phys. Chem. Chem. Phys. {\bf 12}, 11548 (2010).

\bibitem{Barb83} M. N. Barber, in  {\it Phase Transitions and Critical Phenomena Volume 8},
p. 145 (London/New York, Academic Press).

\bibitem{Li92} Y. Li and K. Baberschke, Phys. Rev. Lett. {\bf 68}, 1208 (1992).

\bibitem{Schi96} P. Schilbe, S. Siebentritt, and K.-H. Rieder, Phys. Lett. A {\bf 216}, 20 (1996).

\bibitem{Zhan01} R. Zhang and R. F. Wills, Phys. Rev. Lett. {\bf 86}, 2665 (2001).

\bibitem{Plei04a} M. Pleimling, J. Phys. A: Math. Gen. {\bf 37}, R79 (2004).

\bibitem{Bin00} Ch. Binek, T. Kato, W. Kleemann, O. Petracic, D. Bertrand, F. Bourdarot, P. Burlet,
H. Aruga Katori, K. Katsumata, K. Prokes, and S. Welzel, Eur. Phys. J. B {\bf 15}, 35 (2000).

\end{thebibliography}
\end{document}